\documentstyle[11pt,newpasp,twoside,epsf]{article}
\def\edcomment#1{\iffalse\marginpar{\raggedright\sl#1\/}\else\relax\fi}
\reversemarginpar

\newcommand{\kms}{km~s$^{-1}$}

\newcommand{\lya}{Ly$\alpha$}
\newcommand{\fuse}{{\it FUSE}}
\newcommand{\hst}{{\it HST}}

\begin{document}
\title{A Composite Extreme Ultraviolet QSO Spectrum from FUSE}

\author{Jennifer Scott, Gerard Kriss}
\affil{Space Telescope Science Institute,
3700 San Martin Dr., Baltimore, MD,  21218 USA}
\author{Michael Brotherton}
\affil{University of Wyoming, Department of Physics and Astronomy,
Laramie, WY, 82071 USA}
\author{Richard Green}
\affil{Kitt Peak National Observatory, National Optical Astronomy Observatories,
950 North Cherry Avenue, Tucson, AZ 85726 USA}
\author{John Hutchings}
\affil{Herzberg Institute of Astrophysics, National
Research Council Canada, Victoria, BC V9E 2E7  Canada}
\author{J.\ Michael Shull}
\affil{Center for Astrophysics and Space Astronomy,
Department of Astrophysical and Planetary Sciences, University of Colorado,
Boulder, CO 80309 USA}
\author{Wei Zheng}
\affil{Center for Astrophysical Sciences, Department of Physics and Astronomy,
The Johns Hopkins University, Baltimore, MD 21218 USA}

\begin{abstract}

The Far Ultraviolet Spectroscopic Explorer (\fuse) has surveyed a large
sample ($ > 100$) of active galactic nuclei in the low redshift
universe ($z < 1$).  Its response at short wavelengths
makes it possible to measure directly
the EUV spectral shape of QSOs and Seyfert 1 galaxies at $z < 0.3$.
Using archival
\fuse\ spectra, we form a composite extreme 
ultraviolet (EUV) spectrum of QSOs at $z < 1$ and
compare it to UV/optical composite spectra of QSOs at higher redshift,
particularly the composite spectrum
from archival Hubble Space Telescope spectra.

\end{abstract}

\section{Introduction}

The ubiquity with which QSOs display spectral properties such as 
power-law continua and broad emission lines 
over wide ranges in 
luminosity and redshift has led to the use of composite spectra
to study their global properties.  
Information about the continuum in the rest-frame
ultraviolet is particularly critical
for understanding the formation of the emission lines, for characterizing the Big Blue Bump, 
and for determining the ionization state of the intergalactic medium (IGM).
Composite QSO spectra covering the rest-frame ultraviolet
have been constructed for objects  with
$0.33 < z < 3.6$ from \hst\ (Zheng et al.\ 1997, Telfer et al.\ 2002, T02 hereafter),
and at $z > 2$ from ground-based samples like the
SDSS (Vanden Berk et al.\ 2001),
the First Bright Quasar Survey (Brotherton et al.\ 2001) and the Large Bright Quasar
Survey (Francis et al.\ 1991).

The \fuse\ bandpass,
905-1187 \AA, allows us to
examine the EUV properties of local AGN.  
We can therefore study the same rest-frame wavelength region covered by
the \hst\ composite spectra, at redshifts less than $0.33$.
The low redshifts of these AGN ensure that, although
the \fuse\ aperture limits it to observing relatively bright AGN,
our sample contains a large fraction of intrinsically low-luminosity objects.
An added advantage to working with low redshift spectra
is that the determination of the mean EUV spectral
index requires a less significant correction for IGM absorption than was
required for  the \hst\ sample.

\section{{\it FUSE} Spectra and Composite Construction}

Similarly to T02, we
excluded spectra of broad absorption line quasars and spectra with 
$S/N \la 1$ over large portions
from our \fuse\ sample.
We also exclude spectra of objects that show strong narrow emission lines, 
strong stellar features, or
strong interstellar molecular hydrogen absorption.
A total of 128 spectra of 90 AGN meet the criteria for inclusion in the sample.
We follow the same procedure as T02 for the reduction of
the sample spectra.  To summarize:  
we correct for Galactic extinction using a standard extinction curve, 
individual $E(B-V)$ values for each AGN sightline, 
and $R_{V}=3.1$; we ignore wavelength regions affected by ISM absorption lines;
we correct for Lyman limit absorption 
if the S/N below the Lyman break is greater than one;
we apply a statistical correction 
for the line of sight absorption due to the \lya\ forest; we 
shift the AGN spectrum to the rest frame and resample to common 1 \AA\ bins.

The lower redshifts of the \fuse\ AGN compared with the \hst\ sample of T02 compels 
us to use different 
parameters to perform the correction 
for \lya\ forest absorption mentioned above.
Like T02, we describe the distribution of absorbers
by $\partial^{2}n/ \partial z \partial N \propto (1+z)^{\gamma} N^{-\beta}$.
For the column density distribution parameter, we use the result
found by Dav\'{e} \& Tripp (2001) from 
echelle spectra of
two QSOs at $z\sim0.3$,
$\beta=2.0$ for $12.2 < \log{N} < 14.4$.
For $14.4 < \log{N} < 16.7$, we use $\beta=1.35$ from the 
study by Penton et al.\ (2000).
For the redshift distribution parameter, we use $\gamma=0.15$  (Weymann et al.\ 1998).
We normalize the distribution of absorbers 
by $1.34 \times 10^{-11}$ cm$^{2}$ at $\log{N}=13$ and $z=0.17$ and assume
a Doppler parameter of 21 \kms\ (Dav\'{e} \& Tripp 2001).

\begin{figure}
\plotfiddle{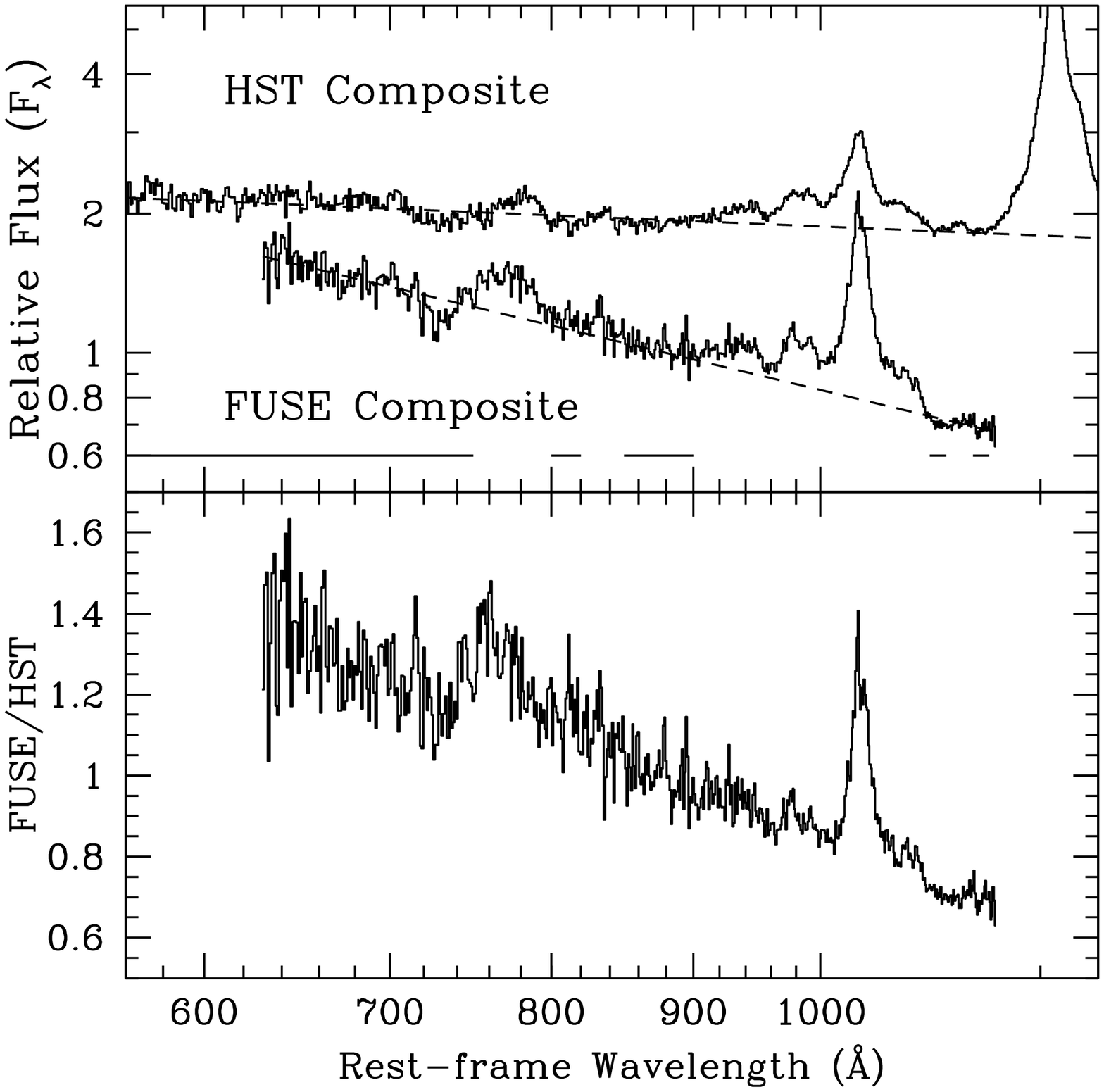}{4.2cm}{0}{30}{30}{-185}{-72}
\plotfiddle{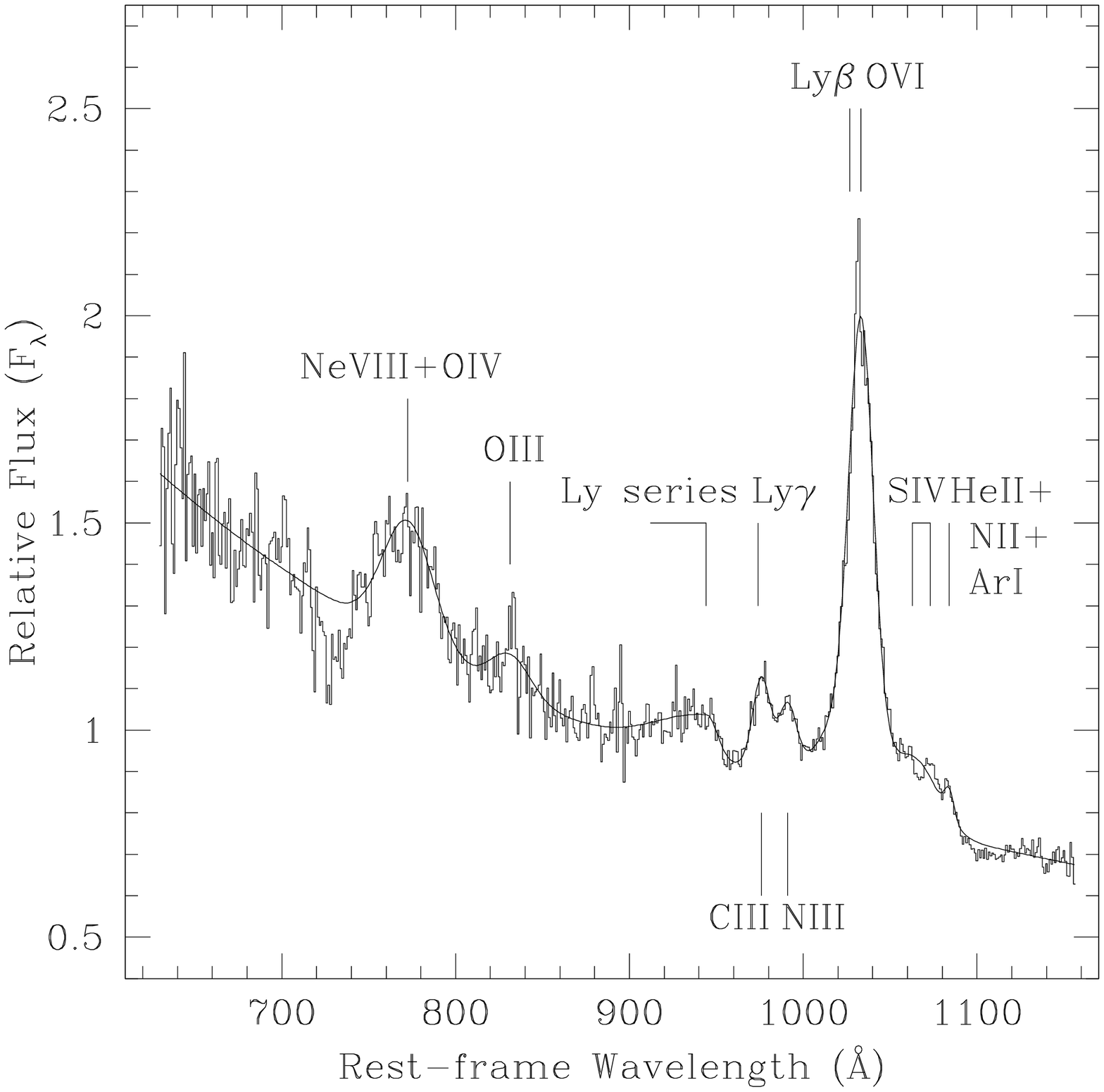}{0cm}{0}{30}{30}{0}{-49}
\caption{{\it Top left:} Composite AGN spectrum with power law 
continuum fit shown by dashed line and
wavelength regions used in fit shown by solid line at bottom;
\hst\ composite from T02 shown for comparison.
{\it Bottom left:}  Ratio of \fuse\ to \hst\ composite spectra.
{\it Right:} Composite AGN spectrum with emission line fit.}
\end{figure}
We combine the sample spectra using the bootstrap technique described by T02.
We begin the bootstrap procedure at the central portion of
the output composite. We then include spectra that fall at longer wavelengths  in
sorted order to longer wavelengths, and spectra at shorter wavelengths
in sorted order to shorter wavelengths.  The overall composite is renormalized at
each step. 
We fit a power law of
the form  $F_{\nu} \propto \nu^{\alpha}$
to the continuum of the composite spectrum; and find that
the best-fit
power law index is $\alpha= -0.56$.  This is
equivalent to the $\alpha_{\rm EUV}$ fits to wavelengths $>500$ \AA\ in T02.
We show the composite spectrum and its ratio to the \hst\ composite in the left
panels of Figure 1.
On the right, we show the emission lines fit to the \fuse\ composite. 

We find a standard deviation of 0.11 in $\alpha$ from 1000 bootstrap samples of the 
\fuse\ sample.
This gives an estimate of the error arising from the range of spectral shapes 
of the individual
AGN that constitute our sample.
We explored a number of possible 
systematic errors that could affect the spectral shape of the \fuse\ composite spectrum.
The results are sensitive to the extinction
correction.  Changing all individual values of $E(B-V)$ by $\pm 1\sigma$, where we
estimate $1\sigma=0.16 E(B-V)$ (Schlegel, Finkbeiner, \& Davis 1998) changes $\alpha$ by $\pm 0.16$.
Changing $R_{V}$ in the extinction law to 4.0(2.8) changes $\alpha$ by $-0.19(+0.06)$.
The composite is also sensitive to the value of the
column density distribution parameter, $\beta$. Reducing it from the chosen
fiducial value of 2.0 to 1.7 or 1.5 increases $\alpha$ by up to 0.3.
We estimate the total error from the items above, 
$\alpha=-0.56^{+0.38}_{-0.28}$.
The EUV spectral index of the \hst\ composite is significantly
softer, $\alpha=-1.76\pm0.12$.
In Figure 2, we show the redshift and luminosity distributions of the \fuse\
and \hst\ samples.

\begin{figure}
\plotfiddle{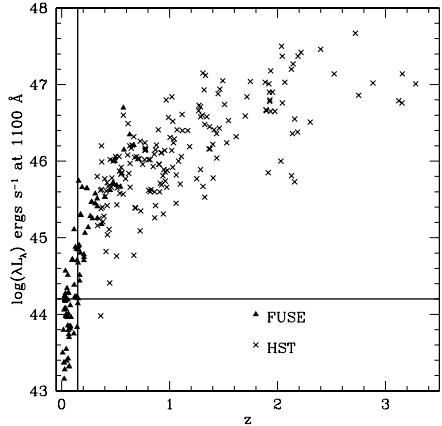}{4.5cm}{0}{30}{30}{-95}{-58}
\caption{Luminosity versus redshift for \fuse\ and \hst\ AGN
with lines marking redshift and luminosity cuts.}
\end{figure}

\section{Summary}

We summarize our results as follows:  

(1) We construct a composite EUV (630-1155 \AA)
AGN spectrum of objects with $z < 0.67$ from archival \fuse\ data.
(2) We fit a power law continuum and Gaussian profiles
to the emission lines in the composite spectrum, and
we find that O~{\sc vi}/Ly$\beta$ and Ne~{\sc viii} emission  are enhanced
in the \fuse\ composite relative to the \hst\ composite due to
the Baldwin effect, seen also in the comparison between
the \hst\ and SDSS composites (T02).
(3) We find that the best-fit spectral index of the composite is
$\alpha=-0.56^{+0.38}_{-0.28}$.
The conservative estimate of the total error in the spectral index
includes the standard deviation in $\alpha$ in
1000 bootstrap samples of the \fuse\ data set,
uncertainties in the extinction correction applied
to the \fuse\ spectra, and uncertainties in the
column density distribution parameter of the intervening Ly$\alpha$ forest.
(4) The \fuse\ composite is 
harder than the EUV portion of the \hst\ composite
spectrum of T02 who find
$\alpha=-1.76 \pm 0.12$ for 332 spectra of 184 AGN with $z>0.33$.
The Baldwin effect is generally attributed to the tendency for
low-luminosity AGN tend to show harder ionizing continua
(Zheng \& Malkan 1993; Wang et al.\ 1998; Dietrich et al.\ 2002).
The median luminosity of the AGN in the \fuse\ sample is
$\log L_{\rm median}=41.2$,
versus $\log L_{\rm median}=42.9$ for the \hst\ sample.
One interpretation of these results is that
both the enhanced high-ionization emission line strengths and
the harder continuum shape of the \fuse\ composite spectrum
are due to the larger fraction
of low-luminosity AGN in the \fuse\ sample.  However, we note
that splitting the \fuse\ 
sample itself
into high- and low-luminosity/redshift
subsamples as shown in Figure 2 results in only 
marginally different values for the EUV spectral index.

\end{document}